\documentclass[preprint,prb,preprintnumbers,amsmath,amssymb]{revtex4}

\usepackage{graphicx}
\usepackage{dcolumn}
\usepackage{bm}
\usepackage{ae}
\usepackage{aecompl}
\usepackage{amsmath}
\usepackage{amssymb}
\usepackage{epsfig}

\begin{document}

\renewcommand{\thefootnote}{\fnsymbol{footnote}}

\title{ The thickness of a liquid layer on the free surface of ice as obtained from computer simulation.
\footnote{J.Chem.Phys. 129 014702 (2008)} }

\author{ M. M. Conde and C. Vega  }
\affiliation{Departamento de Quimica Fisica, Facultad de Ciencias
Quimicas, Universidad Complutense, 28040, Madrid, SPAIN }
\author{  A.  Patrykiejew   }
\affiliation{ Faculty of Chemistry, MCS University, 20031, Lublin, POLAND  }

\date{\today}
\begin{abstract}
Molecular dynamic simulations were performed for
ice $I_{h}$ with a free surface by using four water models, SPC/E, TIP4P, TIP4P/Ice and 
TIP4P/2005. The behavior of the basal plane, the primary prismatic plane and of the
secondary prismatic plane when exposed to vacuum was analyzed. 
We observe the formation of a thin liquid layer at the ice surface at temperatures below the melting point
for all models and the three planes considered.
For a given plane it was found that the thickness of a liquid layer was similar for
different water models, when the comparison is made at the same undercooling with respect to
 the melting point of the model. The liquid layer thickness is found to increase
with temperature. For a fixed temperature it was found that the thickness of the liquid
layer decreases in the following order: the basal plane, the primary prismatic plane, and the secondary prismatic 
plane. For the TIP4P/Ice model, a model reproducing the experimental value of the 
melting temperature of ice, the first clear indication of the formation of a 
liquid layer appears at about 
-100 Celsius for the basal plane, at about -80 Celsius for the primary prismatic plane and at about -70 Celsius
for the secondary prismatic plane.  
\end{abstract}
\maketitle

\section{INTRODUCTION}
The hypothesis of the formation of a liquid-like transition layer on the surface of ice at 
temperatures below the bulk melting point temperature has been the subject of intermittent controversy 
since it was first proposed by Faraday\cite{faraday}. It is now commonly accepted that melting starts at the surface and solids exhibit a liquid-like layer at the surface, already at the temperatures lower than the bulk melting point \cite{abraham1981,Dash_89,Bienfait_92,broughton1983,Nenov_84,Fran_Veen_85,frenken_prb,thompson_surface_melting,mecanismo_melting_LJ}. When the thickness of that liquid layer diverges at the melting point, the melting is denoted as surface melting. When the thickness of the liquid layer remains finite at the melting point this is denoted as         incomplete surface melting \cite{Carnaveli_etal_87,pluis1987,Tartaglino_etal_05}. 
The thickness of a quasi-liquid layer at a given temperature depends on the material considered 
        and on the crystallographic plane exposed (as labelled by the Miller indexes).
        Whenever the thickness of a liquid layer is sufficiently large, either because the 
        system undergoes a surface melting or an incomplete surface melting (with a significantly 
        thick liquid layer), it is not possible to superheat a solid. 
        Due to the ubiquitous character of water, it is of particular interest to 
        determine the structure of the surface of ice\cite{buch_new,kroes_old,pettersson,pettersson2} and, in particular, the structure of a water liquid layer on ice at the temperatures below the melting point. The solution to that problem is not only important from a fundamental point 
        of view, but also from a practical point of view. The existence of 
        a liquid layer of water on the free surface of ice at temperatures below the
        melting point (usually denoted as a quasi-liquid water to illustrate the 
        point that although liquid  in character it is not fully equivalent to bulk water) 
        is relevant to describe different phenomena. For instance it is one of the explanations
        provided to understand why is so easy to skate on ice \cite{physics_today}, although 
        frictional heating is also playing an important role\cite{colbeck1,colbeck2}.
        Also it has been suggested that the existence of  
        a liquid layer on ice, may play an important role in chemical reactions occurring 
        in the stratospheric clouds leading to the annual depletion of ozone in the 
        Antarctic region\cite{molina}. 
        Thus, a number of important  problems are connected  in one way or another 
        to the existence of such a liquid water layer at the ice surface and that explain the appearance of 
        several review papers that have appeared 
        recently\cite{review_somorjai,review_dash,review_dash_2}.

        Is there any evidence of the existence of a quasi-liquid water layer on the free surface
        of ice ? Yes, all experiments point out to 
        the existence of such a quasi-liquid water layer\cite{JCG_1993_129_491,salmeron,JPCS_11_4229_1978,JCP_62_4444_1975,SS_366_43_1996,JCG_82_665_1987,SS_96_357_1980}. 
        Unfortunately there is no consensus 
        about its thickness. In fact, the thickness of such layers
        may differ by an order of magnitude (sometimes almost by two) depending on the 
        technique used\cite{review_somorjai,review_dash}. The problem is certainly difficult 
        and to understand the 
        origin of these discrepancies  further work is still needed. Therefore, it is of interest to 
        see if quasi-liquid layers of water are also observed in 
        computer simulations. Is is the main goal of this work to investigate this problem. 

             When performing computer simulations of ice, it is necessary to choose 
        a model of water. If the goal of the simulations is to establish 
        the existence of a quasi-liquid water layer at temperatures below the melting 
        point, it is absolutely needed to know the melting point of the different 
        water models. Some of the most popular water models, as TIP3P \cite{JCP_1983_79_00926}, 
        TIP4P \cite{JCP_1983_79_00926} and SPC/E \cite{berendsen87}, 
        which are now used in computer simulations on a routine basis were proposed in 
        the early eighties. Somewhat surprisingly the melting points were not established
        on a firm basis, and only the pioneering work 
        of Haymet and co-workers\cite{haymetspcmelting,haymettip4pmelting,JCP_2002_116_08876,bryk02} and 
        Tanaka and co-workers\cite{gao00} provided the first reasonable estimates of the melting point of ice Ih for these models. 
        These early values of the melting temperature were about 10K above the current estimate of the ice
        melting point\cite{sanz1,tanaka04,vegafilosofico,ramon06,wang05,abascal07_jml}. 
        Fortunately, the melting points of TIP4P and SPC/E models have been 
        established on a firm basis over the last three years, being equal to 230(3)K and 215(4)K respectively\cite{sanz1,sanz2,tanaka04,vegafilosofico,ramon06,wang05,abascal07_jml,JCP_2002_116_08876,bryk02,jungwirth_spce}.
        Free energy calculations performed by our group in Madrid\cite{sanz1,sanz2,vegafilosofico} and 
        by Tanaka and co-workers\cite{tanaka04},
        are consistent with the quoted values. Besides, direct simulations of the ice -- liquid water coexistence, done by our 
        group\cite{ramon06}, by Vrba and Jungwirth \cite{jungwirth_spce}, and by Wang et al. \cite{wang05} have also
        confirmed those values of the melting point temperature.
        It is clear that for these two models the melting point is rather low as compared
        to the experimental value. It should be emphasized that not only the melting points but also complete phase diagrams have been determined for these
        models. It has been shown that the TIP4P model is superior
        to SPC/E in describing the global phase diagram of water\cite{sanz1}. For this reason, we have modified
        slightly the parameters of the original TIP4P model to reproduce either the experimental
        value of the melting point, named TIP4P/Ice model,
        or to reproduce the temperature of maximum density of the liquid at the room pressure isobar,
         the model named TIP4P/2005 \cite{abascal05b}.
        The mel\-ting point of TIP4P/Ice and TIP4P/2005 models are of 271(3)K and 249(3)K 
        respectively. Once the melting point of these models is established firmly it seems
        a proper moment to analyze the existence of a quasi-liquid water layer at the tem\-pe\-ra\-tu\-res
        below the melting point. The interest in determining the existence of a quasi-liquid water
        layer is growing enormously in recent years, following the pioneering works of 
        Kroes\cite{kroes92} and of Furukawa and Nada \cite{nada97,nada_furukawa_2}. In fact in the last 
        three years several groups have
        analyzed this issue in detail. Carignano et al. \cite{carignano05} and Ikeda-Fukazawa and Kawamura
        \cite{kawamura04} have clearly shown the existence of 
        the quasi-liquid water layer for two different water models. 
        Carignano et al. \cite{carignano07_ions} have also studied the effect on impurities (NaCl) on the
        quasi-liquid layer using the TIP6P model of water proposed by Nada and van 
        der Eerden\cite{nada03}. Quantum effects on the quasi liquid layer
        have also been considered by Paesani and Voth 
        in a recent study \cite{surface_melting_quantum}. The behavior of a thin film of ice l
        deposited on MgO has been study in detail by Picaud \cite{picaud_jcp06}. 
        Also we have recently reported \cite{maria06} the existence of the liquid layer for TIP4P, 
        SPC/E, TIP4P/Ice and TIP4P/2005 models. In our recent work\cite{maria06,mcbride1}  it has been        shown that superheating
        of ice Ih is suppressed by the existence of a free surface, since the existence of 
        such a free surface induces the formation of a quasi-liquid layer, thus reducing 
        to zero the activation energy of the formation of liquid nucleus\cite{DP_2002_47_00667_nolotengo,MS_2004_30_0397_preprint}, as first speculated 
        by Frenkel \cite{frenkel_kinetics}. Besides we have shown recently that the study of the free surface provides 
        a new methodology to estimate the melting point of water models. In fact the melting 
        point temperature obtained from our simulations of the free surface was fully consistent with 
        the values obtained by other routes and by different authors. If the emphasis of our previous work\cite{maria06} 
        was to propose a new methodology to estimate the melting point of water models, and 
        to show the absence of superheating of ice with a free surface, in this work we 
        focus on the study of quasi-liquid water layer and its thickness, using different water models and different free surfaces of ice.

\section{METHODOLOGY}
 Although the ice $I_{h}$ 
is hexagonal, it is possible to use an orthorhombic unit cell\cite{petrenko99}.
It was with this orthorhombic unit cell that we generated the initial
slab of ice.  In ice $I_{h}$,  protons are 
disordered while still fulfilling the Bernal-Fowler rules 
\cite{bernal33,pauling35}. We used the algorithm of 
Buch et al.\cite{buch98} to obtain 
an initial configuration with proton disorder and  almost zero dipole
moment (less than 0.1 Debye) and satisfying the Bernal Fowler rules.
In order to equilibrate the solid, NpT simulations of bulk ice $I_{h}$ were performed at zero pressure 
for each temperature considered.
We used the molecular dynamics package Gromacs (version 3.3)\cite{gromacs33}.
The time step was 1fs and the geometry
of the water molecules was enforced using constraints\cite{shake,berendsen84}.
The Lennard-Jones potential (LJ) was truncated at 9.0~\AA. 
Ewald sums were used to deal with electrostatics.
The real part of the coulombic potential was truncated at 9.0~\AA.
The Fourier part of the Ewald sums was evaluated by using the Particle Mesh
Ewald (PME) method of Essmann et al.\cite{essmann95}.
The width of the mesh was 1~\AA\ and we used a fourth order
polynomial.  The temperature was kept by using a 
Nos{\'e}-Hoover\cite{nose84,hoover85} thermostat with a relaxation time of 2~ps.
To keep the pressure constant, a Parrinello-Rahman 
barostat\cite{parrinello81,nose83}, with all three sides 
of the simulation box were allowed
to fluctuate independently, was used.  The relaxation time of the barostat was of 2~ps.
The pressure of the barostat was set to zero. 
 The angles were kept orthogonal during this NpT 
run, so that they were not modified with respect to the initial configuration. 
The use of a barostat allowing for independent
fluctuations of the lengths of the simulation box sides is important.
In this way, the solid  can relax to equilibrium by adjusting the
 unit cell size and shape for the considered model and thermodynamic conditions. It is not a good idea to 
impose the geometry of the unit cell. The system should rather determine it from NpT runs. 
Once the ice is equilibrated at zero pressure we proceed to generate the ice-vacuum interface. 
By convention, we shall assume in this paper that the x axis is perpendicular to the
ice-vacuum interface. The ice-vacuum configuration was prepared by simply changing the 
box dimension along the x axis, from the value obtained from the NpT simulations of bulk ice to 
a much larger value. The size of the simulation in the y and z dimension was 
not modified. 
In other words, a slab of ice was located in the middle of a simulation box, and periodic boundary
conditions were used in the three directions of space. The size of the box in the x direction,
was about three times lager than the size of the ice in the y and z directions, so that it is expected that  
the results obtained are not affected by the interaction of the ice with its periodical image. 

Three different planes were used here as the free ice-vacuum interface. First, the se\-con\-da\-ry prismatic
plane (i.e the 1\={2}10 plane) was considered. In this case we used 1024 molecules, and the approximate size of the
simulation box was of 100~\AA\ x 31~\AA\ x  29~\AA\ (the size of the cell in both the y and z directions are only approximate since the actual 
values were obtained from the NpT runs at zero pressure for each water model and tem\-pe\-ra\-tu\-re).
The approximate area of the ice-vacuum 
interface was about 31~\AA\ x 29~\AA\ . This is about 10 molecular diameters in each 
direction parallel to the interface. Although the interface properties present important finite size 
effects\cite{alejandre07}, there is a certain consensus about the fact 
that 10 molecular diameters provide reliable 
estimates of the surface tension of the vapor-liquid interface for water and other systems, and it
seems reasonable to expect that the same is true for the ice-vacuum interface \cite{vega07a}.

We have also analyzed  the behavior of ice, when the plane exposed to vacuum was the 
basal plane. In this case we used 1536 molecules and the dimensions of the simulation box 
were 110~\AA\ x 31~\AA\ x  27~\AA\ (approximately since the y and z value were slightly different
for each water model and temperature). Finally we have also considered the prismatic primary plane.
In this case, we used 1536 molecules and the dimensions of the simulation box were  110~\AA\ x 30~\AA\ x  27~\AA\ .  In Table I, the details of the geometry of the initial configuration are summarized. 

More details regarding the relation of the main planes of 
ice (basal, primary prismatic and secondary prismatic) 
to the hexagonal unit cell can be found in 
figure 1 of the paper by Nada and Furukawa \cite{nada05}, in the paper by 
Carignano et al.\cite{carignano05} and also at the water web site of Chaplin \cite{chaplin}.
Let us just mention that when an orthorhombic unit cell of ice Ih is used, the faces of the
unit cell are just the secondary prismatic plane, the primary prismatic plane and the basal plane. 
Hexagons are clearly visible when the crystal is looked at from the basal plane and from the secondary 
prismatic plane, but not when it is viewed from the side of primary prismatic plane. 
In the basal plane one of the sides of the hexagons 
is parallel to the edge where the basal and the secondary prismatic planes intersect (this is 
not true for the secondary prismatic plane) providing a simple way to distinguish the basal and 
the secondary prismatic planes. 

Once the ice-vacuum system was prepared, we performed relatively long NVT runs (the lengths was 
between 6ns and 12ns  depending on the water model and thermodynamic conditions). Since 
we have been using NVT molecular dynamics, the dimensions of the simulation box 
have been fixed, of course, unlike in the preceding NpT run. 
During this NVT simulations, configurations were stored for further analysis every 4-8ps, after an 
equilibration period of about 2ns. Thus, typically at the end of a run, about 1500 independent 
configurations were available for analysis.

  Once the simulations were performed we could proceed to analyze the configurations obtained.
Since the purpose of this study is to determine the liquid layer thickness at the free surface of 
ice, a criterion allowing to distinguish liquid like and ice like molecules is needed. 
We should admit from the very beginning that there is not a unique procedure
to do that. 
Mapping a configuration (with a continuous
set of coordinates of the molecules) into a discrete set, i.e the numbers of liquid-like and ice-like molecules
requires to establish the borders between the two phases. Also, going from a configuration of water 
molecules (with a continuous set of coordinates of the molecules) to a number of hydrogen bonds 
require a somewhat arbitrary definition of the hydrogen bond. Not surprisingly papers appear 
almost every year providing a new definition of the hydrogen bond, and therefore give different 
 estimations of the hydrogen bonds present in water\cite{matsumoto_hb}.  
Our approach here is to
establish a reasonable criterion to distinguish liquid-like from ice-like 
molecules in a given configuration. We do hope that this allows for a qualitative discussion of 
the results and a progress in the field. We admit that different definitions of ice-like and 
liquid-like molecules may yield different results than reported here, but the qualitative
discussion would remain very much the same. We do really doubt that an absolute, non arbitrary 
definition of what is liquid-like and ice-like part of the system does exist when one deals with a problem in which both phases
are in contact, as it is the case of the free surface of ice.  
To define ice-like and liquid-like molecules we shall use the tetrahedral order parameter 
first introduced by Errington and Debenedetti \cite{errington_nature} and that was found quite 
useful to describe 
the structure of glassy water\cite{glassy_water}. For each molecules ($i$), this parameter $q_{i}$ is defined as:
\begin{equation}
\label{ec1}
q_{i}=[1-\frac{3}{8}\sum^{3}_{j=1}\sum^{4}_{k=j+1}(\cos( \theta_{j,i,k}) +\frac{1}{3})^2]
\end{equation}
where the sum is over the four nearest neighbours (oxygens) of the oxygen of the $i$-th 
water molecule. The angle $\theta_{j,i,k}$ is the angle formed by the oxygens of molecules 
j, i  and k (being molecule i the vertex of the angle). The tetrahedral order parameter adopts a value of 1, when 
the four nearest neighbours adopt a tetrahedral arrangement around the central one.
Notice that negative values of q are also possible when the four nearest neighbours adopt
a linear like configuration.
Let us define p(q), the probability density p(q), as:
\begin{equation}
    p(q)  =    <  \frac{N(q)}{(N \Delta q)}  > 
\end{equation}
where N(q) is the number of molecules  in a given configuration having an order 
parameter between q and $q+\Delta q$, with $\Delta q$ being the size of the grid.
The total number of molecules in the system is denoted as N.  
The brackets in the above equation mean an ensemble average. Obviously the probability 
density p(q) is normalized to unity so that:
\begin{equation}
                1 = \int  p(q)   dq 
\end{equation}

\begin{figure}[ht]
\begin{center}
\caption{
 Probability density distribution $p(q)$ of the orientational order parameter q,
for the water models as obtained at the melting point temperature of
the model (at room pressure) from simulations of bulk water and bulk ice Ih.
(A) TIP4P/Ice; (B) TIP4P/2005; (C) TIP4P and (D) SPC/E.  }
\vspace*{0.5cm}
\includegraphics*[width=85mm,angle=0]{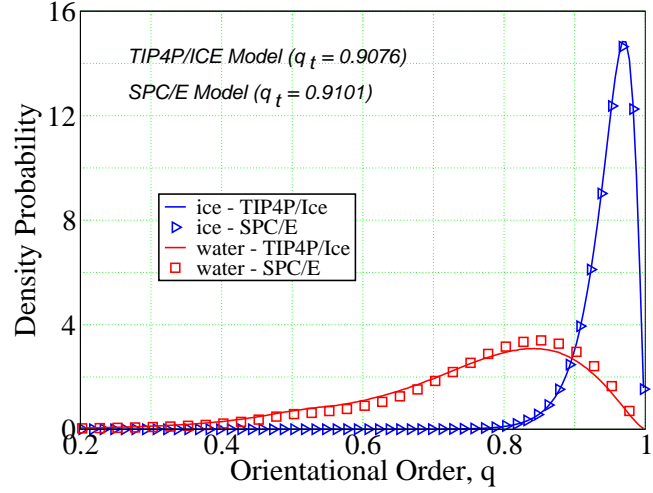} \\
\label{figure1}
\end{center}
\end{figure}

We have analyzed the distribution function of the order parameter q, in pure ice $p_{Ih}(q)$ 
and in pure liquid $p_{liquid}(q)$. For that purpose independent simulations were performed for bulk 
water and bulk ice.
In figure \ref{figure1}, the distributions p(q) are presented for the TIP4P/Ice and SPC/E water models.
The results were obtained at the respective melting point temperatures. 
 As it is seen, the results for these models are quite similar 
(when they are compared at their respective melting point temperatures).
 For the ice Ih the distribution of p(q) is rather narrow and centered around 0.98, whereas
the distribution for water is broader and centered around 0.85. Not surprisingly there is 
a significant amount of tetrahedral order in liquid water at the melting point temperature. 
Certainly our results provide further evidence\cite{headgordon} of the existence of 
tetrahedral order in liquid water in spite of certain claims challenging
this point of view\cite{wernet}. 
The distributions for liquid water obtained by us are in agreement with those presented recently 
by Y. I. Jhon et al.\cite{JPCB_111_9897_2007} for liquid water. The two distributions, $p_{Ih}(q)$ and $p_{liquid}(q)$ 
exhibit a certain degree of overlap . We shall define a threshold value for q, i.e $q_{t}$ so that 
if the value of $q$ of a given molecule is larger than $q_{t}$ the molecule is considered as 
being ice-like , while for the value of $q$ smaller than $q_{t}$ it is considered as a liquid-like.
The threshold value will be obtained for each model from the relation:
\begin{equation}
               \int_{q_t}^{1} p_{liquid}(q) dq  =  \int_{0}^{q_t}  p_{Ih}(q)   dq
\end{equation}
Where $\int_{q_t}^{1}p_{liquid}(q)dq$ is the probability of incorrectly assigning an liquidlike as icelike and $\int_{0}^{q_t}p_{Ih}(q)dq$ is the probability of incorrectly assigning an icelike as liquidlike. In other words, the threshold value $q_t$ is the value of q at which the area of $p_{liquid}(q)$
(for values of q larger than $q_t$ ) is equal to the area under the curve $p_{Ih}(q)$  (for 
values of q smaller than $q_t$).
So, by equating the probability of incorrectly assigning, the errors cancel out. For the threshold value there are as many water molecules having a value of $q$ larger than 
$q_t$ as molecules of ice Ih having a value of $q$ smaller than $q_t$. 
 The threshold values for different water models at the 
melting point temperature are given in Table II. As it is seen the threshold value of q is practically 
identical for all water models. For this reason the value $q_t=0.91$ will be used hereafter 
for all water models. In a given configuration a molecule will be labelled as a liquid-like whenever the 
value of $q$ is smaller than  $q_t=0.91$ and will be classified as an ice-like whenever its 
value of $q$ is larger than $q_t=0.91$. Notice that $q_{t}$ is close to (but not identical with) the point where $p_{liquid}(q)$ and $p_{Ih}(q)$ intersect.

\section{RESULTS AND DISCUSSION}

\begin{figure}[ht]
\begin{center}
\caption{ Instantaneous number of liquid molecules, $N_{liquid}$, as a function of the simulation time
for the TIP4P/Ice model (secondary prismatic plane).}
\vspace*{0.5cm}
\includegraphics*[width=85mm,angle=0]{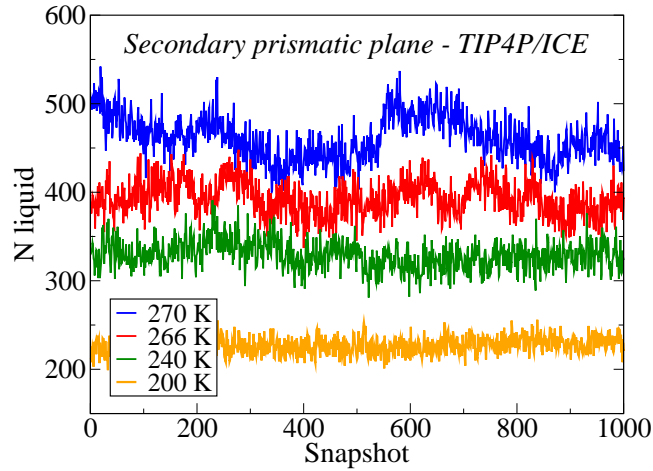} \\
\label{figure2}
\end{center}
\end{figure}

 We shall start with the presentation of results for the TIP4P/Ice model. 
In figure \ref{figure2} the instantaneous number of liquid-like  molecules is presented as a function 
of the simulation time for the TIP4P/Ice model. The results correspond to the case in which 
the secondary prismatic plane is exposed to vacuum. It is quite evident that the number
of liquid-like molecules increases significantly with temperature and
fluctuates around the corresponding average values. The 
fluctuations increase with temperature as well. 

The thickness of the liquid-like layer has been  
estimated as follows:
\begin{equation}
\label{ec5}
   \delta_{apparent} ( \AA\ ) =\frac{N_{liquid}M}{2\rho N_{AV} L_{y}L_{z}10^{-24}}
\end{equation}
where $N_{liquid}$ is the average number of liquid-like molecules along the run, $M$ is the molecular weight of water (18.01574 g/mol), $N_{AV}$ is Avogadro's number, 
the product of $L_y$ and $L_z$ is the area of the interface when both lengths are
given in ~\AA\ , $10^{-24}$ is the factor needed to covert $cm^3$ to \AA\ $^{3}$, while $\rho$ is the 
density of liquid water in $g/cm^3$. The factor of two appears in the denominator of the above equation
 due to the fact that the ice block exhibits two identical interfaces, one appearing on the left size and the other on 
the right side of the simulation cell. As already mentioned, the values of $L_y$ and $L_z$ change 
slightly from one model to another, and from one temperature to another. To compute 
the liquid layer thickness we shall neglect these small variations and we always use
the values reported in Table I. Concerning the density of water, which appear in the
denominator of Eq.\ref{ec5}, it also changes from one water model to another and with the temperature. 
However, for the water models considered in this work the density of water at the melting
point\cite{abascal07_jml} is close to $0.99g/cm^3$. For this reason we shall use this value to estimate the 
liquid layer thickness, $\delta_{apparent}$. The origin of the subscript "apparent" will be
clarified later on.  In summary, the thickness of a liquid layer 
is obtained by equating the average number of liquid-like molecules obtained from 
the simulation of the free interface to that of a sample of pure water of 
dimensions $L_y$ $L_z$  $\delta_{apparent}$.  

\begin{figure}[ht]
\begin{center}
\caption{ Instantaneous values of the liquid layer thickness, $\delta_{apparent}$, as a function of the simulation time for
the TIP4P/Ice model (secondary prismatic plane).}
\vspace*{0.5cm}
\includegraphics*[width=85mm,angle=0]{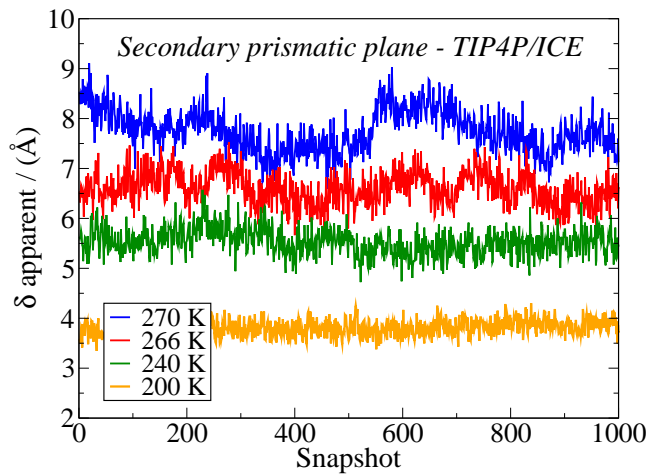} \\
\label{figure3}
\end{center}
\end{figure}

The instantaneous values (along the run) of the liquid layer thickness of the secondary 
prismatic plane of the TIP4P/Ice model are presented in figure \ref{figure3}. 
As it is seen, the fluc\-tua\-tions of about 1~\AA\ in the liquid layer thickness are clearly visible 
along the run at the highest temperature, whereas these fluctuations are much smaller at the 
lowest temperature. 

  In Table \ref{tableIII} the thickness of the liquid layer, $\delta_{apparent}$  as a 
function of temperature is reported 
for the secondary prismatic plane. Results are presented from very low tem\-pe\-ra\-tu\-res 
to temperatures quite close to the melting point. A somewhat surprising result is that
the value of the liquid layer thickness is not zero even at a temperature as low as 30K.
An inspection of snapshots from the simulations at that temperature reveals the absence of a liquid 
layer. One has to ask the equation: why is the number of liquid molecules and hence the thickness of a liquid layer non zero at 30K?  The reason, is that our definition of 
$q$ involves the four nearest neighbours of each molecule. For the molecules occupying the
very last layer of ice it is not possible to form four hydrogen bonds\cite{review_somorjai} 
(so  that at least one 
of the four nearest neighbours is not located in a tetrahedral way).
 Therefore, the value of $q$ for the molecules occupying the last layer 
is rather low and hence such molecules are (incorrectly) classified as liquid-like.
For this reason we found it more convenient to define the true liquid layer thickness as:
\begin{equation}
  \delta_{true} (T)  =  \delta_{apparent} (T)  -  \delta_{apparent} (T=30K) 
 \end{equation}
 with this definition the liquid layer thickness goes to zero at low temperatures 
 at it should.  

\begin{figure}[ht]
\begin{center}
\caption{ The Thickness of a liquid layer for the secondary prismatic plane as a function of the undercooling (A) and of the
 reduced undercooling (B) of the model for the different models of water potentials.
}
\vspace*{0.5cm}
\includegraphics*[width=85mm,angle=0]{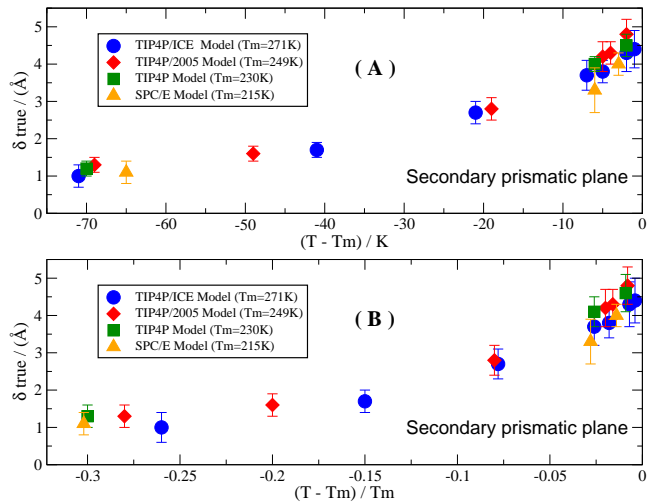} \\
\label{figure4}
\end{center}
\end{figure}

Let us now consider the results for other water models. Table \ref{tableIV} gives the 
values of the liquid layer thickness obtained for the TIP4P/2005, TIP4P, 
and SPC/E models of water.
The results reported were obtained for the secondary prismatic plane, and they
 are similar to those obtained for the TIP4P/Ice model. Notice that the 
temperatures considered for these models are lower than those analyzed for the 
TIP4P/Ice model. The reason is that whereas for TIP4P/Ice the 
melting point reproduces the experimental value, the melting points of TIP4P/2005, TIP4P and SPC/E 
models are about 20K, 40K and 60K below the experimental value, respectively. 
The magnitudes of the liquid layer thickness  obtained for the SPC/E,  
TIP4P and TIP4P/2005 models are similar to those obtained for the TIP4P/Ice. In figure \ref{figure4} the thickness
of the liquid layer at the secondary prismatic plane as a function of the undercooling and of the reduced undercooling is presented for the four water models.  
It is quite well seen that the results of these water models falls into a unique curve, indicating that 
there are no significant differences between the models, when they are compared at 
the same undercooling or reduced undercooling. In summary it is not a good idea
to compare the values of the liquid layer thickness for two different water models, 
at a given absolute temperature.  The liquid layer thickness of two water models should 
be compared at the same undercooling (in degrees) wiht respect to the melting point. The results of 
figure \ref{figure4} manifest clearly that the existence of a liquid layer is not a characteristic 
feature of one specific water model, but is clearly visible in all models considered 
in this work.  Besides, the results strongly suggest that the thickness of the liquid layer is almost 
the same for the different water models, when compared at the same relative temperature to
the melting point. For this reason in what follows we shall consider only the TIP4P/Ice 
model, since it is likely that similar results would be obtained with other water models. 

\begin{figure}[ht]
\begin{center}
\caption{ Instantaneous values of the liquid layer thickness, $\delta_{apparent}$, as a function of the simulation time
for the TIP4P/Ice model (basal plane).}
\vspace*{0.5cm}
\includegraphics*[width=85mm,angle=0]{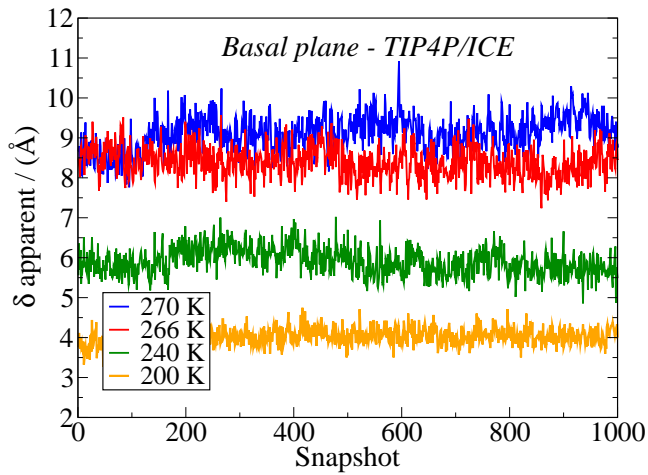} \\
\label{figure5}
\end{center}
\end{figure}

\begin{figure}[ht]
\begin{center}
\caption{ Instantaneous values of the liquid layer thickness, $\delta_{apparent}$, as a function of the simulation time
for the TIP4P/Ice model (primary prismatic plane).}
\vspace*{0.5cm}
\includegraphics*[width=85mm,angle=0]{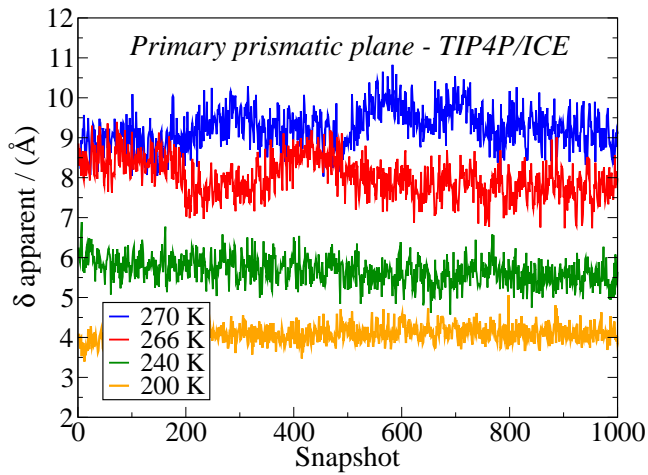} \\
\label{figure6}
\end{center}
\end{figure}

\begin{figure}[ht]
\begin{center}
\caption{ The thickness of a liquid layer for the TIP4P/Ice model at the secondary prismatic, basal and primary prismatic pl
ane as a function
of the undercooling (A) and of the reduced undercooling (B) of the model.
Included are the experimental values by Bluhm et al.\cite{salmeron}}.
\vspace*{0.5cm}
\includegraphics*[width=85mm,angle=0]{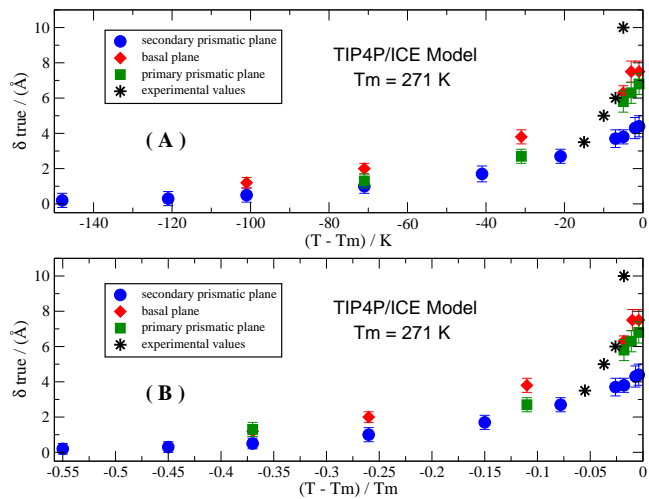} \\
\label{figure7}
\end{center}
\end{figure}

  In figures \ref{figure5} and \ref{figure6} the thickness of a liquid layer 
as obtained along the run are presented, when the plane exposed to vacuum is the 
basal plane and primary prismatic plane, res\-pec\-ti\-ve\-ly. 
The results for the liquid layer thickness (both $\delta_{apparent}$ and $\delta_{true}$)
are summarized in Table \ref{tableIII}. 
The first interesting thing to be noted is that at a given temperature the liquid
layer thickness adopts the largest value for the basal plane and the smallest one for 
the secondary prismatic plane. The value of the primary prismatic plane lies between the
 two. This is more clearly seen in figure \ref{figure7} where the value of $\delta_{true}$ for the 
three planes is plotted as a function of T for the TIP4P/Ice model. 
An interesting question is the following: at which temperature  does the formation of a
liquid layer begins?  We shall denote this temperature as the pre-melting temperature.
Somewhat arbitrarily (again) we shall define the pre-melting temperature as this at which 
$\delta_{true}$ adopts the value of 1~\AA\ . In Table \ref{tableV} the pre-melting temperatures for the
different water models and different ice planes are presented. The data show that the pre-melting 
temperature is located around 70 degrees below the melting point for the secondary 
prismatic plane 
temperature ( this approximately true for all the considered water models). In the case of the primary 
prismatic plane, the pre-melting 
tem\-pe\-ra\-tu\-re is located about 80K below the melting temperature. 
For the basal plane, however, the pre-melting starts at a temperature about 100 degrees below
the melting temperature. It is clear that the basal plane must play a crucial role in the
physics of ice, since it is the plane for which the surface melting starts at the lowest temperature
and it is the plane for which at a given temperature the thickness of the liquid layer is the largest. 

\begin{figure}[ht]
\begin{center}
\caption{ The density profile for the TIP4P/Ice model and the secondary prismatic plane exposed to vacuum.
}
\vspace*{0.5cm}
\includegraphics*[width=85mm,angle=0]{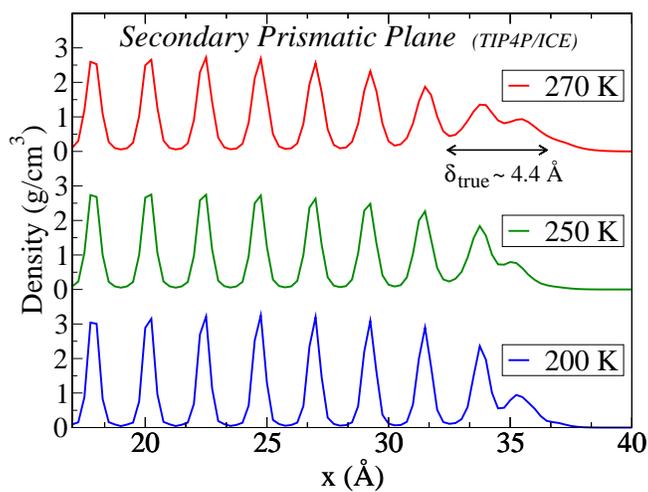} \\
\label{figuredps}
\end{center}
\end{figure}

\begin{figure}[ht]
\begin{center}
\caption{ The density profile for the TIP4P/Ice model and the basal plane exposed to vacuum.
}
\vspace*{0.5cm}
\includegraphics*[width=85mm,angle=0]{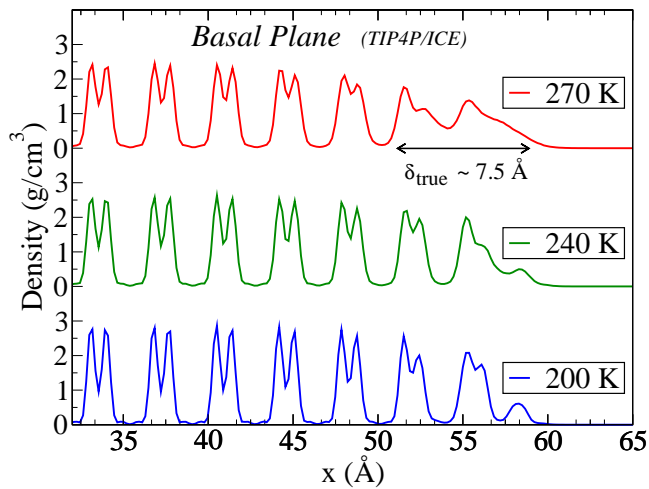} \\
\label{figuredpb}
\end{center}
\end{figure}

\begin{figure}[ht]
\begin{center}
\caption{ Density profile for the TIP4P/Ice model and the primary prismatic plane exposed to vacuum.
}
\vspace*{0.5cm}
\includegraphics*[width=85mm,angle=0]{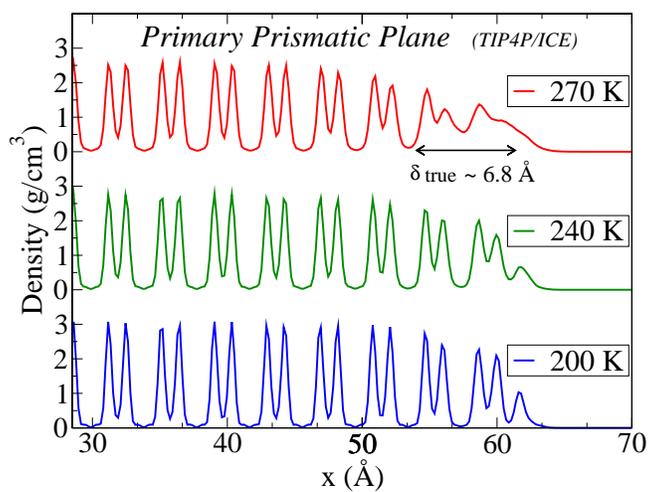} \\
\label{figuredpp}
\end{center}
\end{figure}

 Let us now present the results for density profiles. In figure \ref{figuredps}, \ref{figuredpb} and \ref{figuredpp} the average density profiles for the secondary pris\-ma\-tic plane, the 
basal plane and for the primary prismatic plane at three different temperatures are presented. 
These plots provide a qualitative idea of 
the number of layers involved in the formation of a liquid film. At low temperatures only the
most external layer of water molecules is involved in the formation of a liquid layer. 
However, at temperatures closer to the melting point two additional ice layers appear as 
melted. As it is seen each ice layer of the basal plane has a double peak, the first con\-tai\-ning 
three oxygens of the hexagonal ring and the second peak corresponding to the other three oxygens (located at slightly different values of $x$ coordinate). The same is true for the primary prismatic plane.  In summary the liquid
film involves the first layer of ice at low temperature, and two additional layers at temperatures below (but not too close) to the melting temperatures. 

 It would be of interest to analyze the scaling behavior of the liquid layer thickness in the 
vicinity of the melting point (where a divergence may occur). Although such a study has not been attempted     
here, nevertheless it may be of interest in the future. The reason why the 
study of the divergence may still require further work is twofold. Firstly, it would be necessary 
to determine the melting point of the different water models with still higher accuracy, reducing the estimated uncertainty
by at least one order of magnitude (from 3K of the current estimates to 0.3K). Secondly to analyze the
divergence much larger system sizes are required, so that one has a large piece of ice in contact
with a liquid layer or rather large area. Roughly speaking, systems sizes of about 20000 molecules, simulations 
times of about 100ns, and uncertainty in the melting point temperature of about 0.3K are needed to analyze the possible 
divergence of the liquid layer thickness as the melting point is approached. 

\begin{figure}[ht]
\begin{center}
\caption{ The evolution of the system energy  with time at the
temperatures of $300K$, $290K$ and $276K$ for the TIP4P/Ice model obtained by MD simulation. The plane
exposed to vacuum was the secondary prismatic plane. The final plateau indicates
a complete melting of ice (the temperatures used are above the melting point
of the model). }
\vspace*{0.5cm}
\includegraphics*[width=85mm,angle=0]{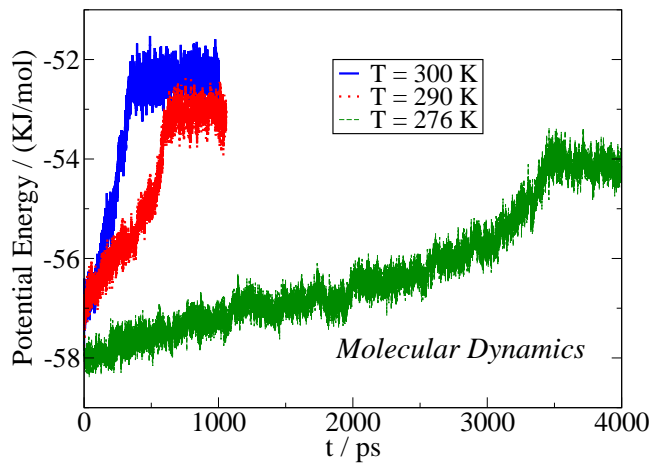} \\
\label{figure11}
\end{center}
\end{figure}

 A different issue is whether the simulation times used in this work (being of the order of 
several ns ) are sufficiently long to guarantee reliable estimates of the liquid layer thickness.
There are two indications suggesting that this is indeed the case. Firstly,  the values 
of the liquid layer thickness exhibit
fluctuations around average values, and the value of the average does not tend to increase or
decrease with time. Secondly, the lengths of simulation runs considered in this work are 
sufficient to melt the block of ice completely, when the temperature is above the melting point. 
In figure \ref{figure11} the evolution of the potential energy of the system with time is plotted at several 
temperatures, namely, 300K, 290K and 276K (for the TIP4P/Ice model the melting point of ice Ih 
 is around 271K). As it can be seen even at the lowest temperature (276K) 
it is possible to melt the ice completely in about $3.5$ ns. 
The final plateau indicates that a complete melting of ice does occur. Obviously, the formation of a liquid layer
with a thickness of about 2 or 3 molecular diameter requires 
less time than a complete melting of the entire ice block.
Although the diffusion coefficient is expected to decrease when the temperature becomes lower, we do not expect 
a dramatic decrease when the temperature drops from say 276K to 266K. For this reason, a relatively fast melting of the ice 
block in about $3.5$ ns at 276K, strongly suggests that a run of similar length should be enough to obtain 
a liquid layer thickness of few ~\AA\ . Another indication that the system is well equilibrated
is the fact that the average number of liquid molecules is the same for the right and the left interfaces.
For the TIP4P/Ice model we have also performed a long run of 30ns of the 
secondary prismatic plane at $T=266K$  and the liquid layer thickness obtained was almost identical to that obtained from a shorter run of about 8ns.

\begin{figure}[ht]
\begin{center}
\caption{ Evolution of the energy of the system with time at $290K$ for the TIP4P/Ice model obtained by MD and MC simulations. 
A single step of MC simulation (a trial move per particle) corresponds to a single time step in MD simulation. }
\vspace*{0.5cm}
\includegraphics*[width=85mm,angle=0]{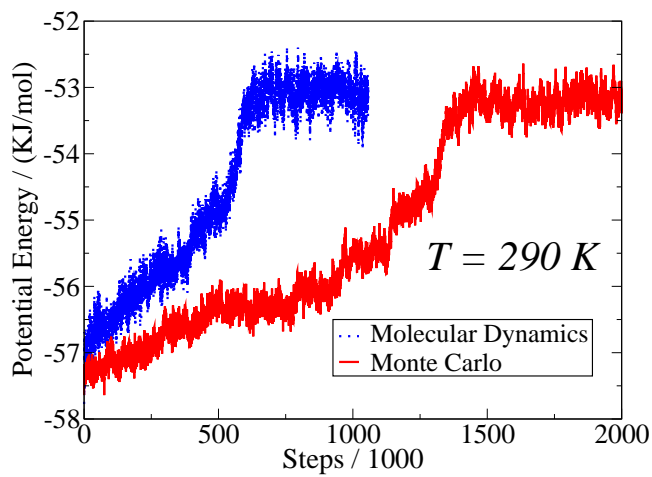} \\
\label{figure12}
\end{center}
\end{figure}

Although we have used MD simulations in our study, nevertheless Monte Carlo si\-mu\-la\-tions could also be used to 
determine the liquid layer thickness. Motivated by this we have performed NVT Monte Carlo simulations
for 
the TIP4P/Ice model at the temperatures of 290K and 300K (both 
above the melting temperature of the model).
Monte Carlo runs were performed using the same initial configuration as used in Molecular Dynamics simulation.
The potential was truncated at the same distance, and rest of the conditions were similar to those
of the Molecular Dynamics simulations (Ewald sums instead of PME were used in this case to deal with 
the long range electrostatics).  
The evolution of the energy of the system with the number of steps is presented in figure \ref{figure12} for $T=290K$. The results are presented 
so that one MD time step corresponds to a Monte Carlo cycle (i.e., a trial move per molecule). 
The figure shows that Monte Carlo simulation requires about 1.4 million cycles to melt the block of 
ice completely, while MD simulations (using a thermostat) required only 0.7 millions to
melt the ice completely. Quite similar results have been obtained at the temperature of 300K. 
 Not only the value of the melting point temperature but also the thickness of the 
liquid layer should be the same, regardless of whether it was obtained via Molecular Dynamics or 
Monte Carlo. 
In fact, by using our Monte Carlo program we have determined the liquid layer thickness for the
TIP4P/Ice model at T=250K  . The length of the run was of 8 million cycles. 
The value of thickness for the secondary prismatic plane obtained ($\delta_{apparent}$) by MC was of  5.7(3)~\AA\ which is in complete agreement 
with the value obtained by MD i.e 5.6(3)~\AA\ .That constitutes a further crosschecking of the results of this work. 

  It was mentioned in the introduction that there is a great disparity in the value of the
liquid layer thickness obtained by different experimental 
techniques \cite{JCG_1993_129_491,salmeron,JPCS_11_4229_1978,JCP_62_4444_1975,SS_366_43_1996,JCG_82_665_1987,SS_96_357_1980}. In figure \ref{figure7} the thickness
of the liquid layer obtained from photoelectron spectroscopy by Salmeron 
and co-workers \cite{salmeron} are presented and compared
with the values obtained in this work (for $\delta_{true}$).
This comparison should be taken with a certain care
since the criterion used to define a liquid layer may be different in experiment and in the simulations
of this work. However, the agreement obtained is reasonable and the simulations seem to describe
the experimental data qualitatively at least. 
It is not obvious at this stage whether
the agreement for the liquid like layer thickness obtained from
computer simulation with that obtained from photoelectron spectroscopy
experiments is accidental or is due to the presence of a common underlying
observable. Further work is required to clarify this point. As it was already 
stated, the system sizes 
considered in this work allow us to obtain reliable values of the liquid 
layer thickness when
it is not larger than about 10-12~\AA\ . When the liquid layer becomes thicker (at the temperatures very close to the melting point)
larger system sizes would be required. 

 In this work we have used a geometrical criterium to identify liquid and solid molecules.
That presents the advantage of simplicity since the geometrical analysis of a certain 
snapshot allows to classify  each molecule as liquid or solid. 
One may wonder whether a dynamic criterium could also be used to identify fluid and solid molecules.
Although that makes the analysis more involved it is worth to explore this possibility. 
Here we shall present some results for the 
secondary prismatic plane of the TIP4P/2005 model. Simulations of bulk ice 
Ih (withouth interfaces) showed clearly that the plateau value of the 
mean square displacement was between $0.1$ \AA $^{2}$ (100K) and  $0.35$ \AA $^{2}$ (250K).
Practically no water molecule presented an individual mean square displacement larger than 
$1$ \AA $^{2}$. On the other hand we found that for TIP4P/2005 (which reproduces reasonably way 
the diffusion coefficient of real water) , the mean square displacement in the 
bulk liquid was larger than 
$1$ \AA $^{2}$ after 400ps (at least for temperatures above 200K). Thus we decided to establish
a simple dynamic criterium to classify  the molecules as liquid or solid.
A molecule will be classified as liquid if after 400ps, its square displacement is 
larger than $1$ \AA $^{2}$ and will be solid otherwise. We do not pretend here to provide a quite elaborate dynamic definition but rather to establish a simple criterium. In Table \ref{tableVI} 
the thickness of the quasi-liquid layer determined by the dynamic criterium are provided.
As it can be seen the thickness of the quasi liquid layer tend to zero at low temperatures.  
The thickness determined from the dynamic criteria is not identical to that
obtained from the geometrical criteria. However they are of the same order of magnitude and 
they agree reasonably well. Thus probaby both geometrical and dynamic criteria can be used to 
identify liquid and solid molecules to determine the liquid layer thickness although in this work we have used extensively the geometric criteria. 

\begin{figure}[ht]
\begin{center}
\caption{ Instantaneous configuration of the TIP4P/Ice system at 268K at the end of a 5 ns run. Although the temperature is well below the melting point of the model, a quasi-liquid layer is clearly present at the ice-vacuum interface. The plane that can be seen is the secondary prismatic plane. The plane exposed to the vacuum is the basal plane.}
\vspace*{0.5cm}
\includegraphics*[width=85mm,angle=0]{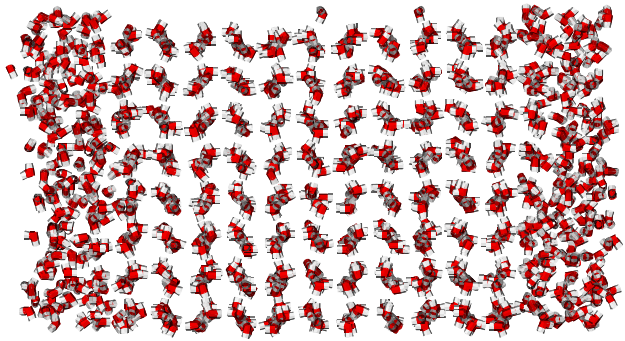} \\
\label{figure13}
\end{center}
\end{figure}

  Finally since a picture is worth a 
thousand words, let us finish by presenting an ins\-tan\-ta\-neous
configuration obtained for the TIP4P/Ice model at T=268K in the basal plane. This is presented in figure \ref{figure13}. The
picture shows a graphical evidence of the existence of a liquid layer at the free surface of ice. 

\section{Conclusions}
  In this work we have reported on the results of computer simulation study of the for\-ma\-tion of a liquid layer at the ice surface at the temperatures below the melting point. A slab of ice has been located in the middle of a large simulation box,
and long MD runs have been performed, for different planes of ice exposed to vacuum. Three different 
planes considered: the secondary prismatic plane, the basal plane and the primary prismatic 
plane.  A tetrahedral orientational order parameter has been used to classify each water molecule within 
an instantaneous configuration as being liquid-like or solid-like. 
 When the orientational order parameter was larger than the 
threshold value, set as equal to  $q_{t}=0.91$, the molecule was regarded as an ice-like. When the 
orientational order was smaller then the molecule was classified as a liquid-like. 
In this way the average number of liquid-like molecules was calculated, and
 the liquid layer thickness was estimated. Main findings of this work are as follows:
\begin{itemize}
\item{ There is a clear evidence that a liquid layer develops at the 
free surface of ice at the temperatures below the melting point. }
\item { The appearance of liquid layers starts at different temperatures depending on the ice surface plane exposed to vacuum. It appears at the temperature of about -100 (with respect
to the melting point) for the basal plane, at about -80 for the primary prismatic plane and at about -70 for the
secondary prismatic plane. The thickness of a liquid layer increases with temperature. }
\item{ At a given temperature the thickness of the liquid layer is larger for the basal plane 
than for the primary prismatic plane, and for the primary prismatic plane is larger than for the
secondary prismatic plane. }
\item{ When the thickness of liquid layer for  different water models are compared at the 
same degree of undercooling , then the value of the liquid layer thickness
is practically the same for all water models. } 
\item { The thickness of the liquid layer seems to be of the order of about 10~\AA\ at the temperatures up to 
        3-4K below the melting point. To determine the thickness at temperatures closer to the 
        melting point, larger simulation cells and more accurate estimates of the melting point are needed. }
\item { The thickness of liquid layers determined in this work seem to be of the same order of magnitude as  determined
         by  H. Bluhm et al.\cite{salmeron}, from photoelectron spectroscopy. }
\end{itemize}

\acknowledgements
This work was funded by grants FIS2007-66079-C02-01
from the DGI (Spain), S-0505/ESP/0229 from the CAM, MTKD-CT-2004-509249
from the European Union and 910570 from the UCM.
M.M.Conde would like to thank Universidad Complutense by the award of a 
PhD grant. 

\bibliographystyle{./apsrev}

\newpage
\begin{table}
\label{tableI}
\caption{ Details of the geometry of the initial configuration for different planes of ice corresponding to different water models. $L_x$, $L_y$, $L_z$ are the dimensions of the simulation box 
in ~\AA\ . The x axis is perpendicular to the ice-vacuum interface. $L_{ice}$ is the size of the initial block of ice (in the x direction). All dimensions are given in ~\AA\ .}
\begin{tabular}{cccccc}
\hline
N  &  The plane exposed  &  $L_x$ &  $L_y$  &  $L_z$  & $L_{ice}$  \\
\hline
1024 & Secondary prismatic & 100 & 31 & 29 & 36 \\ 
1536 & Basal               & 110 & 31 & 27 & 59 \\
1536 & Primary prismatic   & 110 & 30 & 27 & 62 \\
\hline
\end{tabular}
\end{table}

\begin{table}
\caption{ Threshold value of the orientational order parameter, $q_t$, for the
different water models (as determined at their respective melting temperatures). }
\label{tableII}
\begin{tabular}{cc}
\hline
Model &
$q_t$ \\
\hline
TIP4P/Ice & 0.9076 \\
TIP4P/2005 & 0.9085 \\
TIP4P & 0.9105 \\
SPC/E & 0.9101 \\
\hline
\end{tabular}
\end{table}

\clearpage

\newpage
\begin{table}
\caption{ Thickness of the liquid layer for different planes of the TIP4P/Ice model. $\delta_{apparent}$ and $\delta_{true}$ are given in ~\AA\ and the temperature is given in K.}
\label{tableIII}
\begin{tabular}{cccc}
\hline
Water model and plane &
T &
$\delta_{apparent}$&
$\delta_{true}$ \\
\hline
&270&7.3(5)&4.4(7)    \\
&269& 7.2(5)&4.3(7)  \\
&266& 6.7(3)&3.8(5)\\
&264&  6.6(4)&3.7(6) \\
TIP4P/Ice & 250& 5.6(3)&2.7(5) \\
(Secondary prism.) & 230&  4.6(2)&1.7(4) \\
&200&  3.9(3)&1.0(5) \\
&170&  3.4(2)&0.5(4) \\
&150&  3.2(2)&0.3(4) \\
&123& 3.1(2)&0.2(4) \\
&30&  2.9(2)& 0.0(2) \\
\hline
&270& 9.6(5)&7.5(7) \\
&268&  9.6(5)&7.5(7)  \\
TIP4P/Ice & 266& 8.4(2)&6.3(4) \\
(Basal) & 240& 5.9(3)&3.8(5) \\
&200& 4.1(2)&2.0(4)\\
&170 & 3.3(2) & 1.2(4) \\
&30&  2.1(2)& 0.0(2) \\
\hline
&270 & 9.5(6)&6.8(8) \\
&268& 9.0(6)&6.3(8) \\
TIP4P/Ice & 266& 8.5(6)&5.8(8)   \\
(Primary prism.) & 240& 5.4(3) &2.7(5)  \\
&200& 4.0(2)  &1.3(4) \\
&30& 2.7(2)  & 0.0(2) \\
\hline
\end{tabular}
\end{table}

\clearpage

\newpage
\begin{table}
\caption{ Thickness of the liquid layer for the different water models (TIP4P/2005, TIP4P and SPC/E). $\delta_{apparent}$ and $\delta_{true}$ are given in \AA\ and the temperature is given in K.}
\label{tableIV}
\begin{tabular}{cccc}
\hline
Water model and plane &
T &
$\delta_{apparent}$&
$\delta_{true}$ \\
\hline
&249&  8.5(1.0)&  6.0(1.2) \\
&247&  7.3(4)&  4.8(6) \\
&245&  6.8(3)&  4.3(5) \\
&244&  6.7(4)&  4.2(6) \\
TIP4P/2005 & 230&  5.3(3)&  2.8(5) \\
(Secondary prism.) & 200&  4.1(2)&  1.6(4) \\
&180&  3.8(2)&  1.3(4) \\
&140&  3.4(2)&  0.9(4) \\
&100&  3.0(2)&  0.5(4) \\
&30&   2.5(2)&  0.0(2) \\
\hline
&228& 7.1(4)&  4.6(6) \\
TIP4P & 224& 6.6(3)&  4.1(5) \\
(Secondary prism.) & 200 & 3.9(2) & 1.4(4) \\
&160&  3.8(2)&  1.3(4) \\
&30&  2.5(2)&  0.0(2) \\
\hline
SPC/E & 212& 7.0(2)&  4.0(4) \\
(Secondary prism.) & 209& 6.3(5)&  3.3(7)  \\
&150& 4.1(2)&  1.1(4)\\
&30 & 3.0(2)&  0.0(2) \\
\hline
\end{tabular}
\end{table}

\clearpage

\newpage
\begin{table}[!hbt]\centering
\caption{ Pre-melting temperature for different water models and different planes of ice. The temperature is given in K. $T_{pre-melting}$ is defined as the temperature at which $\delta_{true}$ is 1 \AA\ .}
\label{tableV}
\begin{tabular}{cccc}
\hline
Model&
$T_{melting}$ &
$T_{pre-melting}$ & 
$T_{pre-melting}$-$T_{melting}$\\
\hline
TIP4P/Ice basal plane &271 (3)&170 & $\sim$ -100   \\
TIP4P/Ice primary prismatic plane & 271 (3)& 190 & $\sim$ -80\\
TIP4P/Ice secondary prismatic plane & 271 (3)& 200 & $\sim$ -70\\
TIP4P/2005 secondary prismatic plane & 249 (3)& 180 & $\sim$ -70\\
TIP4P secondary prismatic plane&230 (3)&160  & $\sim$ -70\\
SPC/E secondary prismatic plane& 215 (4) & 150 & $\sim$ -65\\
\hline
\end{tabular}
\end{table}

\begin{table}
\caption{ Thickness of the liquid layer for TIP4P/2005 as obtained from the 
dynamic criteria $\delta_{dynamic}$ and from the geometric criteria 
$\delta_{apparent}$ and $\delta_{true}$. 
Results were obtained for the secondary prismatic plane. $\delta_{dynamic}$, 
$\delta_{apparent}$ and $\delta_{true}$ are given in \AA\ and 
the temperature is given in K.} 
\label{tableVI}
\begin{tabular}{ccccc}
\hline
Water model and plane &
T  &
$\delta_{dynamic}$ & 
$\delta_{apparent}$ &
$\delta_{true}$ \\
\hline
&249&    8.9(7)& 8.5(1.0) &6.0(1.2)\\
&247&    7.5(5)&7.3(4) &4.8(6)\\
TIP4P/2005 & 230&    3.9(3) &5.3(3) &2.8(5)\\
(Secondary prism.) & 200&    1.5(3) &4.1(2) &1.6(4)\\
&180&    1.2(6) & 3.8(2)&1.3(4)\\
&140&    0.5(1) &3.4(2) &0.9(4)\\
&100&   0.1(1) &3.0(2) &0.5(4)\\
&30&     0.0(1) &2.5(2) &0.0(2)\\
\hline
\hline
\end{tabular}
\end{table}

\clearpage


\clearpage
\newpage

\end{document}